\documentclass{ws-procs9x6}

\begin{document}

\title{Describing the Baryon Spectrum with $1 \! /N_{\rm c}$ QCD}

\author{Richard F. Lebed}

\address{Department of Physics \& Astronomy,\\
Arizona State University,\\ 
Tempe, AZ 85287-1504, USA\\ 
E-mail: richard.lebed@asu.edu}

\maketitle

\abstracts{This talk outlines recent advances using QCD in the
$1/N_c$ limit aimed at understanding baryon scattering processes and
their embedded short-lived baryon resonances.  In this presentation we
emphasize developing qualitative physical insight over presenting
results of detailed calculations.}

\section{Introduction}

When addressing an audience of baryon resonance experts, it is hardly
necessary to emphasize the elusive nature of the $N^*$s as both
experimental and theoretical objects: Owing to their extremely short
$O(10^{-23} \, {\rm s})$ lifetimes, they are often barely discernable,
lurking in baryon scattering amplitudes like strangers in a fog.  My
previous talk write-ups on this material\cite{confs} have been geared
exclusively towards theory audiences, but an $N^*$ conference is
attended by a large number of experimentalists as well, who view
theory talks with an eye toward picking up new notions of physical
understanding for the phenomena that they study, rather than focusing
on calculational detail.  I therefore wish to focus here on the
qualitative description of the motivation behind and the results of my
recent work with Tom Cohen on excited
baryons.\cite{CL1,CLcompat,CDLN1,CLpent,CDLN2,CLSU3,CDLM,CLSU3phenom}
The reader who craves more detail is welcomed to peruse
Refs.~\refcite{confs} or the original works.

\section{Two Physical Pictures for $N^*$s}

The most frequently invoked picture for baryons is that suggested by
the constituent quark model, in which the light (masses $\sim$ 5~MeV)
fundamental quarks of the QCD Lagrangian somehow agglomerate with the
multitude of gluons and virtual quark-antiquark pairs to form
constituent ($\sim$ 300~MeV) quarks.  In order to be discernable as
distinct entities, such pseudoparticles must nevertheless remain
weakly bound to one another.  If this physical picture is valid, then
the baryon, originally a complicated many-body object only describable
using full quantum field theory, reduces to a simple three-particle
quantum-mechanical system interacting through a potential, not unlike
a miniature atomic nucleus.  In this case the baryon excited states
consist of orbital and radial excitations of the three constituents.
Inasmuch as the constituent quark masses are larger than the energies
that bind them, the baryons fill well-defined multiplets based upon
approximate invariances of the state under quark spin flips, quark
flavor substitutions, and spatial exchanges, the SU(6)$\times$O(3)
symmetry.

Constituent quark models therefore predict numerous excited hadron
multiplets, the lowest of which have indeed been observed.  For
example, the ground states, consisting of the nucleons, the $\Delta$
resonances (related to the nucleons by a spin flip), and their strange
partners, fill a spin-flavor-space symmetric $({\bf 56}, 0^+)$ of
SU(6)$\times$O(3), while the lightest excitations appear to fill the
orbitally-excited mixed-symmetry multiplet $({\bf 70}, 1^-)$ or a
radially-excited $({\bf 56}, 0^+)$.  However, higher in the spectrum
the picture becomes much murkier, with numerous partly-filled
multiplets as well as predicted multiplets whose members remain
unobserved.

Alternately, the chiral soliton picture for baryons, starting directly
from a hadronic perspective, recognizes that hadrons rather than
quarks are the states observed in nature.  {\it Solitons\/} are
semiclassical finite-energy solutions to a field theory, which is to
say that they are non-dissipating ``lumps'' of energy (such as a lump
in a rug placed in a room too small: It can be moved from place to
place, but not eliminated).  Chiral Lagrangians, which have been so
successful in delimiting light meson dynamics, admit solitonic
solutions that couple to mesons according to chiral symmetry
constraints.  Their semiclassical nature is guaranteed if they are
heavy compared to the mesons, just as is physically true for the
baryons.  In the best-studied variant, the Skyrme model, the solitons
are shown to carry fermionic statistics.

The basic soliton configuration, called a {\it hedgehog}, turns out
not to possess a single well-defined isospin or spin quantum number,
but rather a quantum number that is the magnitude of their vector sum
${\bf K} \! \equiv \!  {\bf I} \! + \! {\bf J}$, sometimes called the
{\it grand spin}.  Physical baryon states with particular spin and
isospin eigenvalues are then recovered by forming a judicious linear
combination of hedgehog states of different $K$; these ``judicious''
couplings are none other than Clebsch-Gordan coefficients (CGC).  The
couplings of mesons to the underlying hedgehog, as arise in scattering
processes, also induce spin and isospin CGC.

Excited baryons in chiral soliton models appear as rotational or
vibrational excitations of the basic hedgehog configuration.  Much of
the particular spectrum generated by such excitations depends strongly
upon the details of the dynamical ``profile'' functions multiplying
the hedgehog, making predictions of baryon resonance multiplets in
chiral soliton models less than robust.

\section{Large $N_c$ QCD and the $1/N_c$ Expansion}

While both the constituent quark and chiral soliton models warrant
attention for incorporating observable features of baryons, they
remain just that---models.  In both cases, an expansive literature
demonstrates that one may refine the models by including subleading
effects, but it is not {\it a priori\/} obvious which corrections are
essential for understanding baryon dynamics.  Instead, we prefer to
obtain a method directly from QCD that combines the best features of
both pictures.  {\it Ab initio\/} lattice calculations applied to
excited baryons hold great promise for the future,\cite{Morn} but even
when completed will provide numerical results rather than definitive
dynamical statements.

Large $N_c$ QCD, obtained by supposing that QCD had not 3 but some
larger number $N_c$ of color charges, is not a model but rather an
extension of the field theory representing strong interactions.  It is
physically useful if $i$) physical observables have well-defined
limits as $N_c \! \to \! \infty$ [{\it i.e.}, with small $O(1/N_c)$
corrections], and $ii$) the values of these observables do not change
excessively as $N_c$ is allowed to decrease from a large value down to
3\@.  The key question then becomes whether one can recognize in
observables unambiguous signatures of this expansion in powers of
$1/3$, and in fact the $({\bf 56}, 0^+)$ baryons provide ample
evidence\cite{NStar02} in their spectra and couplings.

We first require a few fundamental baryon results.  For $N_c$ colors,
the baryons contain at least $N_c$ quarks, the number required to form
a colorless state.  Baryons have $O(N_c^1)$ masses, and meson
couplings that are $O(N_c^{1/2})$ (trilinear) and $O(N_c^0)$
(quartic).\cite{Witten} The latter fact implies that ordinary baryon
resonances, since they appear in baryon-meson scattering amplitudes,
have masses above the ground states and widths each of $O(N_c^0)$.
The baryons themselves, despite having large masses at large $N_c$,
maintain an essentially constant [$O(N_c^0)$] size, which follows from
the suppression of multiple-quark interactions by powers of $N_c$.
Lastly, order-by-order unitarity in $N_c$ powers in baryon-meson
scattering processes (called {\it consistency
conditions\/}\cite{DM,DJM}) require the ground-state multiplet to have
not only spin-$\frac 1 2$ but spin-$\frac 3 2$ members as well, the
large $N_c$ analogue to the {\bf 56} [for $N_c \! > \! 3$ the
completely symmetric SU(6) multiplet also contains up to
spin-$\frac{N_c}{2}$ states].

Both the quark model and the chiral soliton model have straightforward
extensions to arbitrary $N_c$.  Of course, $N_c$ must be odd for
baryons to remain fermions.  In the quark model case, one may
define\cite{BL} quantum fields with all the properties of constituent
quarks by noting that ground-state baryons carry precisely the quantum
numbers of $N_c$ quarks (which remains true for $N_c \! = \! 3$; this
of course was the original motivation of the quark model), and
dividing the baryon into $N_c$ non-overlapping ``interpolating
fields'' that exhaust its wave function.  Using this definition for
the quarks, the suppression of multiquark operators by powers of
$1/N_c$ allows one to conclude that effects carrying the spin-flavor
quantum numbers of such operators are also suppressed.  If the states
are stable against strong decays (as is the case for the ground-state
multiplet), one may construct a Hamiltonian for which these baryons
are the asymptotic states, and matrix elements are computed by means
of the Wigner-Eckart theorem.  For example, the nucleon and $\Delta$
masses are split only at $O(1/N_c)$ because this is the order of the
lowest-order (hyperfine) Hamiltonian operator distinguishing their
masses; the exact coefficient remains incalculable unless the strong
interactions can be solved from first principles, but if the $1/N_c$
expansion is valid, then it should be a typical hadronic scale (a few
hundred MeV) times an $O(1)$ number.  Indeed, the observed
$N$-$\Delta$ splitting follows this pattern.\cite{Jenk}

One may attempt an extension of this approach to the excited baryons.
A large body of literature\cite{old70} treats (for example) the
lightest negative-parity resonances as filling the analogue to the
$({\bf 70}, 1^-)$, a symmetrized core of $N_c \! - \!  1$ quarks and
one excited quark.  While this approach has yielded many interesting
phenomenological insights, its strict application seems sensible only
when $i$) the excited baryons are also asymptotically stable states of
a Hamiltonian, and $ii$) can be represented uniquely as 1-quark
excitations of a ground state ({\it i.e.}, {\it configuration
mixing\/} with states having 2 or more excited quarks but the same
overall quantum numbers are ignored).

Chiral soliton models also combine efficiently with the $1/N_c$
expansion.  Indeed, much of the interest in such models during the
early 1980s centered on the fact that the semiclassical nature of the
solitons was consistent with the heaviness of large $N_c$ baryons, in
that many of their predictions turned out to be independent of the
particular choice of profile function.\cite{ANW} Subsequent
work\cite{Manohar} showed that quark and soliton models for
ground-state baryons share common group-theoretical features in the
large $N_c$ limit.  But these results apply only to the ground-state
multiplet, whose members are related by various rotations of the basic
hedgehog state.

\section{Resonances in the $1/N_c$ Expansion}

Since soliton models can be used to study baryon scattering
amplitudes, it begs the question whether one can use these models to
reach beyond the ground states and obtain definite statements about
resonances with a degree of model independence inherited from large
$N_c$.  A successful picture for resonances ought not put them in by
hand; they are intrinsically excitations in baryon scattering
amplitudes and should be generated as complex-valued poles ($z_R \! =
\! M_R \! + \! \frac{i}{2}\Gamma_R$) within them.  Work along these
lines in the mid-1980s began with Ref.~\refcite{HEHW} and rapidly
progressed to focus upon model-independent group-theoretical
features:\cite{Mattis} In particular, from this approach one finds a
number of linear relations between distinct partial-wave amplitudes.

The central feature driving these works is the underlying conservation
of $K$-spin.  As we have seen, not only the composition of baryon
states from the hedgehog, but also the couplings of baryon-meson
scattering processes, introduce group-theoretical factors.  Carefully
combining them yields the full set of baryon partial wave amplitudes
written as linear combinations of a smaller set of underlying {\it
reduced amplitudes\/} labeled by $K$, while composing the CGC leads to
coefficients that are purely group-theoretical $6j$ and $9j$ factors.
As a trivial example, for $\pi N$ scattering one obtains $S_{11} \! =
\! S_{31}$.

Based upon interesting regularities noted for scattering processes
viewed in the $t$-exchange channel,\cite{Donohue} $K$-spin
conservation (expressed in terms of the usual $s$-channel quantum
numbers) was shown\cite{MM} to be equivalent to the $t$-channel rule
$I_t \! = \!  J_t$.  It was not until several years later, however,
that the $I_t \! = \!  J_t$ rule was shown\cite{KapSavMan} to follow
directly from large $N_c$ consistency conditions, completing the
ingredients of the proof\cite{CL1} that underlying $K$-spin
conservation is a direct result of the large $N_c$ limit.

To say that full baryon partial waves are linearly related for large
$N_c$ means that a resonant pole occurring in any one of them must
appear in at least one of the others, or more fundamentally, in one of
the reduced amplitudes.  However, since a given reduced amplitude
contributes to multiple partial waves, the same resonant pole appears
in each one: Large $N_c$ baryon resonances appear in multiplets
degenerate in both mass and width.\cite{CL1}

Large $N_c$ baryon resonances are not the exclusive provenance of
soliton models; if one considers the large $N_c$ generalization of the
$({\bf 70}, 1^-)$ using the Hamiltonian approach described above, one
finds\cite{CL1,CLSU3phenom,PS} that only 5 distinct mass eigenvalues
occur up to $O(N_c^0)$ inclusive, the level at which distinct
resonances of the ground states split in mass.  When one examines all
partial waves in which states carrying these quantum numbers can
occur, one finds that all of the states in the multiplet are induced
by one pole in each of the reduced amplitudes with $K \! = \! 0$,
$\frac 1 2$, 1, $\frac 3 2$, and 2 (and only $K \! = \! 0,1,2$ occur
for the nonstrange states).  From the point of view of large $N_c$,
the irreducible multiplet $({\bf 70}, 1^-)$ of SU(6)$\times$O(3) is
therefore actually a reducible collection of 5 {\it distinct\/}
irreducible multiplets, which are labeled by $K \! =
\! 0$, $\frac 1 2$, 1, $\frac 3 2$, and 2; let us label the masses as
$m_K$.  When SU(3) flavor symmetry is invoked, $K$ may also be defined
for strange states, where it is simply defined as the magnitude of
${\bf I} \! + \! {\bf J}$ for the nonstrange member of the SU(3)
multiplet.  A similar pattern, which we call {\it
compatibility},\cite{CLcompat,CLSU3} occurs for every
SU(6)$\times$O(3) multiplet, each of which decomposes at large $N_c$
into a collection of irreducible multiplets labeled by $K$: Each
quark-model multiplet forms a collection of distinct resonance
multiplets.  This result generalizes the one discussed above, that the
ground-state multiplet in large $N_c$ forms a complete $({\bf 56},
0^+)$ (in this case, only $K \! = \! 0$ appears).

\section{Phenomenological Consequences}

The quark and chiral soliton approaches thus find common ground for
large $N_c$ by having compatible resonance multiplets.  But this is a
formal result; to find phenomenological successes, one needs to go no
further than examining which reduced amplitudes appear in a given
partial wave amplitude.

To illustrate this point, let us consider the lightest $I \!  = \!
\frac 1 2$, $J \! = \! \frac 1 2$ ($N_{1/2}$) negative-parity states.
It turns out for any $N_c \! \ge \! 3$ that $({\bf 70}, 1^-)$ contains
precisely 2 $N_{1/2}$ states; for $N_c \! = \! 3$ these are $N(1535)$
and $N(1650)$.  Using only the group theory imposed by the $N_c \! \to
\! \infty$ limit, $\eta N$ states at large $N_c$ allow only $K \! = \!
0$ amplitudes, while the process $\pi N \! \to \! \pi N$ allows only
$K \! = \! 1$.  Thus, only the resonance of mass $m_0$ appears in
$\eta N$ amplitudes, and only $m_1$ appears in $\pi N \! \to \! \pi
N$.  As is well known to this audience, $N(1535)$ lies just barely
above the $\eta N$ threshold and yet decays to it as frequently as to
the heavily phase-space favored $\pi N$ channel.  Alternately, the
$N(1650)$ has a $\pi N$ branching ratio many times larger than for
$\eta N$ despite a much more comparable phase space in these
channels.\cite{CL1} The $N(1535)$ $\pi N$ and $N(1650)$ $\eta N$
couplings thus arise only through subleading corrections of the size
expected from the $1/N_c$ expansion.

Results of this sort also appear among the strange
resonances.\cite{CLSU3phenom} In particular, the $N(1535)$ appears to
be just the nonstrange member of an entire $K \! = \! 0$ octet of
resonances, all of which therefore are $\eta$-philic and $\pi$-phobic.
As evidence, note that the $\Lambda (1670)$ lies only 5~MeV above
$\eta \Lambda (1116)$ threshold, and yet this channel has a 10--25\%
branching ratio.

Even stranger selection rules occur when full SU(3) group theory is
taken into account.\cite{CLSU3phenom} For example, one can prove for
$N_c$ arbitrary that resonances in SU(3) multiplets whose highest
hypercharge states are nonstrange ({\bf 8} and {\bf 10}) decay
preferentially (by a factor $N_c^1$) with a $\pi$ or $\eta$, while
those whose top states are strange ({\bf 1}) prefer $\overline{K}$
decays by $O(N_c^1)$.  Evidence for this peculiar prediction is borne
out by the $\Lambda (1520)$: Its branching ratios for $\overline{K} N$
and $\Sigma \pi$ are roughly equal, but when the near-threshold
$p^{2L+1}$ behavior for this $d$ wave is taken into account, one finds
the effective coupling constant ratio $g(\Lambda(1520) \! \to \!
\overline{K} N)/g(\Lambda(1520) \! \to \! \Sigma \pi) \! \sim 4$--$5
\! = O(N_c)$, as advertised.

$1/N_c$ corrections may be incorporated by noting the demonstration
that the $I_t \! = \! J_t$ rule is equivalent to the large $N_c$
limit\cite{KapSavMan} also shows amplitudes with $|I_t \! - \! J_t| \!
= \!  n$ to be suppressed by at least $1/N_c^n$.  To incorporate all
possible $O(1/N_c)$ effects one simply appends to all possible
amplitudes with $I_t \! = \! J_t$ those with $I_t \! - \! J_t \! = \!
\pm 1$.\cite{CDLN2} The number of reduced amplitudes then increases
while the number of observable partial waves of course remains the
same, making linear relations tougher to obtain; for example, no such
$1/N_c$-corrected relations occur among $\pi N \! \to \! \pi N$, but
$\pi N \! \to \pi \Delta$ relations do occur, and definitely improve
by about a factor of 3 when the $1/N_c$ corrections are taken into
account.\cite{CDLN2}

We have noted that configuration mixing between different states with
the same overall quantum numbers can be a nuisance within specific
models by requiring additional assumptions.  A true advantage of
treating excited baryons as resonances in partial wave amplitudes is
that configuration mixing can occur naturally.  As an example of this
philosophy, if one model predicts an especially narrow excited baryon
[say, a width of $O(1/N_c)$], and if there exist broad resonances
[$O(N_c^0)$] in the same mass region with the same overall quantum
numbers, then generically the states mix and produce two broad
resonances.\cite{CDLN1} In the quark picture, for example, this mixing
occurs any time one can find a Hamiltonian operator with transition
matrix elements of $O(N_c^0)$ between the two states.

The existence of well-defined multiplets of resonances at large $N_c$
is also an aid to searching for exotic states.\cite{CLpent} For
example, let us suppose that the pentaquark candidate $\Theta^+
(1540)$ were confirmed with hypercharge $+2$, $I \! = \! 0$, $J \! =
\! \frac 1 2$, and either parity.  Then large $N_c$, independently of
any model, mandates that it must have $I \! = \! 1$, $J \! = \!
\frac{1}{2} , \frac{3}{2}$ and $I \! = \! 2$, $J \! = \! \frac 3 2$
partners with the same mass [up to $O(1/N_c)$ corrections, less than
about 200~MeV] and the same width [which of course can magnify or
shrink in response to nearby thresholds, again indicating $O(1/N_c)$
differences].

Studies of baryon scattering amplitudes are not limited only to
couplings with mesons.  As long as the quantum numbers and the $1/N_c$
couplings of the field to the baryons is known, precisely the same
methods apply.  Processes such as photoproduction, electroproduction,
real or virtual Compton scattering are then open to scrutiny.  In the
case of pion photoproduction, the photon carries both isovector and
isoscalar quantum numbers, with the former dominating\cite{JJM} by a
factor $N_c$.  Including the leading and first subleading isovector
and the leading isoscalar amplitudes then gives linear relations among
multipole amplitudes with relative $O(1/N_c^2)$
corrections.\cite{CDLM} Some of the relations obtained this way ({\it
e.g.}, the prediction that isovector amplitude combinations dominate
isoscalar ones) agree quite impressively with data.  Some, however, do
not appear to the eye to fare as well.  In those cases, the threshold
behaviors still agree quite well, followed by seemingly disparate
behavior in the respective resonant regions.  Does this mean that the
$1/N_c$ expansion is failing?  Not so: The disagreements come from
resonances in the different partial waves whose masses are split at
$O(1/N_c)$, giving critical behavior occurring in different places in
distinct partial waves.  When this effect is taken into account by
extracting couplings {\em on resonance\/} (as presented by the
Particle Data Group\cite{PDG}), the linear relations good to
$O(1/N_c^2)$ do indeed produce results that agree to within
10--15\%.\cite{CDLM}

\section{Looking Ahead}

A very brief summary tells us where this program is at the current
time: We now have at our disposal the correct large $N_c$ method of
studying baryon resonances of finite widths model-independently, {\it
i.e.}, in the context of a full quantum field theory.  Multiplets of
resonances degenerate in masses and widths naturally arise in this
approach, and are similar but not identical to old quark-model
multiplets.  The first phenomenological results have been very
encouraging, demonstrating that the $1/N_c$ expansion continues to
bear a rich harvest for the excited states.  Not only the resonances
themselves, but the partial wave amplitudes in which they appear, can
be studied using the same methods.

The most important issue yet unsolved in this program is how to treat
spurious states, {\it i.e.}, those that occur only for $N_c \! > \!
3$.  Indeed, we were loose in our notation when we spoke of, for
example, the SU(6) {\bf 56} or the SU(3) {\bf 8}, which contain (due
to quark combinatorics) many more than the given number of states when
$N_c \! > \! 3$.  As commented above, we obtain interesting results
for specific states occurring with the same multiplicities for all
$N_c \! \ge \!  3$, such as negative-parity $N_{1/2}$'s.  However,
many more high-spin and high-isospin states occur for $N_c \! > \! 3$.
Which ones survive at $N_c \! = \! 3$ and which ones do not?  Since
this is a difference between $N_c \! \to \! \infty$ and $N_c \! = \!
3$, it represents a special kind of $1/N_c$ correction yet to be
mastered.

All results thus far obtain from 2-to-2-particle scattering processes.
In fact, multiparticle processes such at $\pi N \! \to \! \pi \pi N$
are not substantially more difficult in many cases of interest.  For
example, if the $\pi \pi$ pair is identified by reconstruction as
originating from a $\rho$, then the suppressed width [$O(1/N_c)$] of
$\rho$ allows the process to be studied in factorized form.

The reader should note that physical input within this method has been
virtually nil: Only the imposition of an organizing principle, around
suppressions in powers of $1/N_c$, has occurred.  In this sense, the
$1/N_c$ methods employed thus far have the flavor of chiral
Lagrangians, which obtain results using only symmetries and a
low-momentum expansion.  Indeed, one thrust of future work will be the
folding of chiral symmetry ({\it e.g.}, low-energy theorems) into the
$1/N_c$ expansion; our preliminary examination suggests this to be a
promising direction.

The essential tools thus appear to be in place to disentangle the
fundamental features of the $N^*$ spectrum using a systematic
approach, much as chiral Lagrangians have done for the light
mesons.  Given sufficient time and resources, it is a program well
within the reach of the $N^*$ community.

\section*{Acknowledgments}
This work was supported in part by the National Science Foundation
under Grant No.\ PHY-0456520.


\begin{thebibliography}{0}

\bibitem{confs}
R.F.~Lebed, {\tt hep-ph/0406236}, published in {\it Continuous
Advances in QCD 2004}, edited by T.~Gherghetta, World Scientific,
Singapore, 2004;
{\tt hep-ph/0501021}, published in {\it Large $N_c$ QCD
2004}, edited by J.L.~Goity {\it et al.}, World Scientific,
Hackensack, NJ, USA (2005);
{\tt hep-ph/0509020}, invited talk at {\it International
Conference on QCD and Hadronic Physics}, Beijing, 16--20 June 2005 (to
appear in proceedings).

\bibitem{CL1}
T.D.~Cohen and R.F.~Lebed, {\it Phys.\ Rev.\ Lett.} {\bf 91}, 012001
(2003); {\it Phys.\ Rev.} D {\bf 67}, 012001 (2003).

\bibitem{CLcompat}
T.D.~Cohen and R.F.~Lebed, {\it Phys.\ Rev.} D {\bf 68}, 056003 (2003).

\bibitem{CDLN1}
T.D.~Cohen, D.C.~Dakin, A.~Nellore, and R.F.~Lebed, {\it Phys.\ Rev.}
D {\bf 69}, 056001 (2004).

\bibitem{CLpent}
T.D.~Cohen and R.F.~Lebed, {\it Phys.\ Lett.} B {\bf 578}, 150 (2004);
{\it Phys.\ Lett.} B {\bf 619}, 115 (2005).

\bibitem{CDLN2}
T.D.~Cohen, D.C.~Dakin, A.~Nellore, and R.F.~Lebed, {\it Phys.\ Rev.}
D {\bf 70}, 056004 (2004).

\bibitem{CLSU3}
T.D.~Cohen and R.F.~Lebed, {\it Phys.\ Rev.} D {\bf 70}, 096015
(2004).

\bibitem{CDLM}
T.D.~Cohen, D.C.~Dakin, R.F.~Lebed, and D.R.~Martin, {\it Phys.\ Rev.}
D {\bf 71}, 076010 (2005).

\bibitem{CLSU3phenom}
T.D.~Cohen and R.F.~Lebed, {\it Phys.\ Rev.} D {\bf 72}, 056001
(2005).

\bibitem{CHL}
T.D.~Cohen, P.M.~Hohler, and R.F.~Lebed, {\it Phys.\ Rev.} D {\bf 72},
074010 (2005).

\bibitem{Morn}
See C.~Morningstar, these proceedings.

\bibitem{NStar02}
R.F.~Lebed, {\tt hep-ph/0301279}, in {\it NStar 2002: Proceedings of
the Workshop on the Physics of Excited Nucleons}, edited by S.A.~Dytman
and E.S.~Swanson, World Scientific, Singapore, 2003.

\bibitem{Witten}
E.~Witten, {\it Nucl.\ Phys.} {\bf B160}, 57 (1979).

\bibitem{DM}
R.F.~Dashen and A.V.~Manohar, {\it Phys.\ Lett.} B {\bf 315}, 425
(1993).

\bibitem{DJM}
R.F.~Dashen, E. Jenkins, and A.V.~Manohar, {\it Phys.\ Rev.} D {\bf
49}, 4713 (1994).

\bibitem{BL}
A.J.~Buchmann and R.F.~Lebed, {\it Phys.\ Rev.} D {\bf 62}, 096005
(2000).

\bibitem{Jenk}
E.~Jenkins, {\it Phys.\ Lett.} B {\bf 315}, 441 (1993); E.~Jenkins and
R.F.~Lebed, {\it Phys.\ Rev.} D {\bf 52}, 282 (1995).

\bibitem{old70}
C.D.~Carone, H.~Georgi, L.~Kaplan, and D.~Morin, {\it Phys.\ Rev.}
D {\bf 50}, 5793 (1994);
J.L.~Goity, {\it Phys.\ Lett.} B {\bf 414}, 140 (1997);
D.~Pirjol and T.-M.~Yan, {\it Phys.\ Rev.} D {\bf 57}, 1449 (1998);
D {\bf 57}, 5434 (1998);
C.E.~Carlson, C.D.~Carone, J.L.~Goity, and R.F.~Lebed, {\it Phys.\
Lett.} B {\bf 438}, 327 (1998); {\it Phys.\ Rev.} D {\bf 59}, 114008
(1999);
J.L.~Goity, C.~Schat, and N.~Scoccola, {\it Phys.\ Rev.\ Lett.} {\bf
88}, 102002 (2002); {\it Phys.\ Rev.} D {\bf 66}, 114014 (2002); {\it
Phys.\ Lett.} B {\bf 564}, 83 (2003);
J.L.~Goity and N.N.~Scoccola, {\it Phys.\ Rev.} D {\bf 72}, 034024
(2005); J.L.~Goity, {\tt hep-ph/0504121};
C.E.~Carlson and C.D.~Carone, {\it Phys.\ Rev.} D {\bf 58}, 053005
(1998); {\it Phys.\ Lett.} B {\bf 441}, 363 (1998); B {\bf 484}, 260
(2000);
N.~Matagne and Fl.~Stancu, {\it Phys.\ Rev.} D {\bf 71}, 014010
(2005); {\it Phys.\ Lett.} B {\bf 631}, 7 (2005); N.~Matagne, {\tt
hep-ph/0501123}; {\tt hep-ph/0511247} (these proceedings).

\bibitem{ANW}
E.~Witten, {\it Nucl.\ Phys.} {\bf B223}, 433 (1983); G.S.~Adkins,
C.R.~Nappi, and E.~Witten, {\it Nucl.\ Phys.} {\bf B228}, 552 (1983);
G.S.~Adkins and C.R.~Nappi, {\it Nucl.\ Phys.} {\bf B249}, 507 (1985).

\bibitem{Manohar}
A.V.~Manohar, {\it Nucl.\ Phys.} {\bf B248}, 19 (1984).

\bibitem{HEHW} 
A.~Hayashi, G.~Eckart, G.~Holzwarth, H.~Walliser, {\it Phys.\ Lett.}
B {\bf 147}, 5 (1984).

\bibitem{Mattis}
M.P.~Mattis and M.~Karliner, {\it Phys.\ Rev.} D {\bf 31}, 2833 (1985);
M.P.~Mattis and M.E.~Peskin, {\it Phys.\ Rev.} D {\bf 32}, 58 (1985);
M.P.~Mattis, {\it Phys.\ Rev.\ Lett.} {\bf 56}, 1103 (1986); {\it
Phys.\ Rev.} D {\bf 39}, 994 (1989); {\it Phys.\ Rev.\ Lett.} {\bf
63}, 1455 (1989);
%
{\it Phys.\ Rev.\ Lett.} {\bf 56}, 1103 (1986).

\bibitem{Donohue}
J.T.~Donohue, {\it Phys.\ Rev.\ Lett.} {\bf 58}, 3 (1987); {\it Phys.\
Rev.} D {\bf 37}, 631 (1988).

\bibitem{MM}
M.P.~Mattis and M.~Mukerjee, {\it Phys.\ Rev.\ Lett.} {\bf 61}, 1344
(1988).

\bibitem{KapSavMan}
D.B.~Kaplan and M.J.~Savage, {\it Phys.\ Lett.} B {\bf 365}, 244
(1996); D.B.~Kaplan and A.V.~Manohar, {\it Phys.\ Rev.} C {\bf 56}, 76
(1997).


\bibitem{PS}
D.~Pirjol and C.~Schat, {\it Phys.\ Rev.} D {\bf 67}, 096009 (2003).


\bibitem{JJM}
E.~Jenkins, X.~Ji, and A.V.~Manohar, {\it Phys.\ Rev.\ Lett.} {\bf
89}, 242001 (2002).

\bibitem{PDG}
{\it Review of Particle Properties} (S.~Eidelman {\it et al.}), {\it
Phys.\ Lett.} B {\bf 592}, 1 (2004).

\end{thebibliography}
\end{document}